\documentclass[twocolumn,showpacs,preprintnumbers,amsmath,amssymb]{revtex4}
\usepackage{graphicx}% Include figure files
\usepackage{dcolumn}% Align table columns on decimal point
\usepackage{bm}% bold math

\newcommand{\ba}{\begin{eqnarray}}
\newcommand{\ea}{\end{eqnarray}}
\newcommand{\bse}{\begin{subequations}}
\newcommand{\ese}{\end{subequations}}
\newcommand{\dd}{\hbox{d}}
\newcommand{\Teq}{T_{\hbox{eq}}}
\newcommand{\M}{{\cal {M}}}
\newcommand{\A}{{\cal {A}}}

\newcommand{\CS}{{\cal {S}}}
\newcommand{\CP}{{\cal {P}}}

\newcommand{\kB}{k_{_B}}

\newcommand{\Vrot}{V_{\textrm{\tiny{rot}}}}
\newcommand{\rhocr}{\rho_{\textrm{\tiny{crit}}}}

\begin{document}

\title{The Maxwell--Boltzmann gas with
non-standard self--interactions: a novel approach to galactic dark
matter}
\author{Alexander Balakin$^*$, Roberto A.
Sussman$^\ddagger$ and Winfried Zimdahl$^\dagger$}
\affiliation{
$^*$ Department of General Relativity and Gravitation,
\\ Kazan State University, \\ 420008 Kazan, Russia.\\ $^\ddagger$
Instituto de Ciencias Nucleares,\\  Apartado Postal 70543, UNAM, \\M\'exico DF,
04510, M\'exico.\\
$^\dagger$ Institut f\"ur Theoretische Physik, \\Universit\"at zu K\"oln,\\
D-50937 K\"oln, Germany.\\}

\begin{abstract}
Using relativistic kinetic theory, we study spherically symmetric,
static equilibrium configurations of a collisionless
Maxwell-Boltzmann gas with non-standard self-interactions,
modelled by an effective one--particle force. The resulting set of
equilibrium conditions represents a generalization of the
classical Tolman-Oppenheimer-Volkov equations.  We specify these
conditions for two types of Lorentz--like forces: one coupled to
the 4-acceleration and the 4--velocity and the other one coupled
to the Riemann tensor. We investigate the weak field limits in
each case and show that they lead to various Newtonian type
configurations that are different from the usual isothermal sphere
characterizing the conventional Newtonian Maxwell--Boltzmann gas.
These configurations could provide a plausible phenomenological
and theoretical description of galactic dark matter halo
structures. We show how the self--interaction may act
phenomenologically as an effective cosmological constant and
discuss possible connections with Modified Newtonian Dynamics
(MOND).
\end{abstract}


\maketitle

\section{Introduction}

There is a broad consensus that both on cosmological and on
galactic scales substantial amounts of dark matter are needed to
explain current observations. Recent data from type Ia supernovae,
the large-scale-structure analysis, and the anisotropies of the
cosmic microwave background (CMB) suggest that the dynamics of the
present Universe is determined by a mixture of roughly 70$\%$ dark
energy and roughly 30\% (cold) dark matter (CDM). To understand
the physical nature of these main ingredients of the cosmic
substratum is one of the basic challenges in current cosmology.
Numerous models have been worked out over the past few years, the
most popular one still being the $\Lambda$CDM model which also
plays the role of a reference setting for competing scenarios,
such as the different quintessence models, which are based on the
dynamics of a scalar field with a suitable
potential~\cite{Qmatter,Qmatter2}. In some models interacting
quintessence played a crucial role~\cite{intQ1,intQ2}.
Furthermore, the cosmic medium has also been modelled as a gas
with non-standard interactions given in terms of an effective
one-particle force~\cite{gasQ1,gasQ2,gasQ3,gasQ4}.

Regarding dark matter (DM), the CDM paradigm of collisionless (yet
undetected) supersymmetric particles is successful at the
cosmological scale~\cite{Ellis,Fornengo}, though some of its
predictions at the galactic scale (``cuspy'' density profiles and
excess substructure) have been at odds with
observations~\cite{cdm_problems_1,cdm_problems_2}. It is natural
then to look for DM models with some type of phenomenological
self--interaction~\cite{scdm}, but because of the unknown nature
of DM in galactic halos, the type of self-interaction is also an
open question. Hence, other models (``warm'' DM) consider lighter
more thermal particles~\cite{wdm} and even non--thermal sources,
such as coherent scalar fields have been
contemplated~\cite{sfdm1,sfdm2,sfdm3,sfdm4}.

Since DM in galactic halos constitutes about $90-95\,\%$ of the
mass of galactic systems, while the latter have Newtonian
characteristic velocities, the dynamics of these systems is
usually examined within a Newtonian framework in which, as a first
approximation, visible matter can be considered as test particles
in the gravitational field of the DM halo. The standard models are
either idealized configurations of collisionless matter obtained
from Newtonian kinetic theory~\cite{BT}, or models based on
n--body numerical simulations~\cite{nbody_1,nbody_2,nbody_3} that
yield ``universal'' density or rotation velocity
profiles~\cite{urc,DMprofs1,DMprofs2,DMprofs3}. While the usual
Newtonian approach may well be adequate on galactic scales, only a
general relativistic setting can provide limits for the validity
of the Newtonian approximation. In a general relativistic
treatment in which DM provides the bulk of spacetime curvature,
the motion of baryonic matter can be treated as test particle
motion in the geometry generated by the DM (``rotation
velocities'' become then velocities of circular geodesic
orbits~\cite{HGDM, Lake}). However, a relativistic approach will
not only yield corrections to the Newtonian case, but it also
provides a framework which admits theoretical generalizations,
e.g. the introduction of non-standard interactions in a systematic
and self--consistent way. It is this property, which will be
relevant for the present paper. Given the unknown nature of the
halo matter, it seems to be of interest to consider various types
of non--standard interactions within this matter and to explore
the consequences of such an assumption. Matter with
self-interaction may have a Newtonian type limit which differs
from the standard Newtonian limit for non-interacting dark matter.
This property is demonstrated here for two choices of a
self-interaction. On this basis it becomes possible to discuss
modifications of standard astrophysical configurations where the
focus of the present paper is on isothermal spheres.

Gas models allow interactions to be described by effective
one-particle forces. To obtain candidates for potentially
interesting self-interactions we follow a strategy similar to the
one which was previously used in a cosmological
context~\cite{gasQ1,gasQ2,gasQ3}. These investigations
demonstrated that the admissible structure of the relevant forces
is severely restricted by the symmetries of the problem (which are
the symmetries of the cosmological principle
in~\cite{gasQ1,gasQ2,gasQ3}). In the present paper we apply this
formalism for the first time to a galactic scale in which
astrophysical systems (DM halos) are necessarily inhomogeneous,
hence we consider a spherically symmetric and static geometry. The
latter is then combined with the equilibrium conditions of a
Maxwell--Boltzmann (MB) gas in the presence of a self-interacting
force field. On the fluid level this force generates effective
pressures, representing additional degrees of freedom that extend
the classical Tolman--Oppenheimer--Volkov (TOV) analysis. We
demonstrate that the resulting configurations provide
modifications of the usual isothermal sphere (the Newtonian limit
of a ``standard'' MB gas), which have the potential to provide
adequate phenomenological and theoretical descriptions of galactic
DM. As a particular feature, the interaction, introduced here in a
general relativistic framework, takes the form of an additional
Newtonian type force in the corresponding limit. This indicates
possible relations to Modified Newtonian Dynamics
(MOND)~\cite{MOND} although, different from the latter, the
interaction here is an interaction within the DM.

The present paper also provides a generalization of previous well
established work on relativistic kinetic theory of collisionless
matter applied to globular clusters~\cite{prevRKT} (though our
focus here is on DM halos).

The paper is organized as follows: sections II and III present the
general formalism of relativistic kinetic theory with
self--interactions modelled as a one particle force. In sections
IV and V we obtain the full general relativistic equilibrium
equations for a static, spherically symmetric spacetime. We
specialize in section VI to a Lorentz--like force, linear in the
particle momenta and without anisotropic stresses. In section VII
we consider two possible cases of covariant couplings of this
Lorentz--like force: firstly, to the 4--acceleration and the
4--velocity and secondly, to the Riemann tensor. Section VIII
investigates two types of Newtonian type limits: a ``standard''
Newtonian limit that yields modified isothermal configurations and
a Newtonian type limit in which the generalized force acts as an
effective cosmological constant. Two numerical examples of
configurations constructed with each type of Newtonian limit are
discussed in section IX, while section X provides a summary and
conclusion.

\section{Kinetic Theory}

We assume that the particles of a collisionless  relativistic gas
move  under the influence of a 4-force field $F^a$. The equations
of motion of the gas particles are~\cite{kinetic}
  \begin{equation} p^a \ = \ m\,\frac{\dd x^a}{\dd\tau},\qquad F^a \ = \
\frac{\hbox{D}p^a}{\dd\tau},\label{pF_def}  \end{equation}
where $\tau$ is a parameter along the particle worldlines. Since
the particle 4-momenta are normalized according to $p_a\,p^a = -
m^2 c^{2}$ with constant mass $m$, the force $F^a$ must satisfy
$p_a\,F^a=0 $. The corresponding equation for the invariant
one-particle distribution function $f=f(x,p)$ may be written as
  \begin{equation} \ p^a\,\frac{\partial f}{\partial
x^a}-\Gamma^a\,_{bc}\,p^b\,p^c\,\frac{\partial f}{\partial
p^a}+m\,\frac{\partial (f\,F^a)}{\partial p^a} \ = \ 0\
.\label{Liou_eq}  \end{equation}
We shall restrict ourselves to the class of forces which admit
solutions of (\ref{Liou_eq}) that are of the type of the
J{\"u}ttner distribution function
  \begin{equation} f\,^0 \ = \ \exp\,(\alpha + \beta_a\,p^a),
  \label{f0_def}  \end{equation}
where $\alpha=\alpha(x)$ and $\beta_a$ is a timelike four vector.
The particle number flow 4-vector, $N^a$, and the equilibrium
energy-momentum tensor, $\Teq^{ab}$, are then defined in a
standard way (see, e.g., \cite{kinetic}) as
\bse\label{NT_def}\ba N^a && = \ c\,\int{\dd P\,p^a\,f^{0}(x,p)},\\
\Teq^{ab} && = \ c\,\int{\dd P\,p^a\,p^b\,f^{0}(x,p)}.  \ea\ese
Here, the integrals are over the mass shell $m =$ const in
momentum space. For the balance equations we find
\ba N^a_{\ ;a} && = \ -mc\int{\dd P\frac{\partial
(f^{0}\,F^a)}{\partial p^a}} \ = \ 0\ ,\label{Nbal_def}\\
\Teq^{ab}\,_{;b} && = -m\ c\ \int{\dd P\,p^a\,\frac{\partial
(f^{0}\,F^b)}{\partial p^b}}\nonumber \\ &&= \ m\,c\,\int{\dd
P\,f^{0}\,F^a}\ ,\label{Tbal_def}\ea
where the semicolon denotes the covariant derivative.

\section{Symmetry considerations}

The quantities $N^a$ and $\Teq^{ab}$ may be split with respect to
the unique 4-velocity $u^a$ as
  \begin{equation} N^a \ = \ n\,u^a,\qquad \Teq^{ab} \ = \
  \frac{E}{c^{2}}\,u^a\,u^b +
P\,h^{ab},\label{NTf0_def}  \end{equation}
where $h^{ab}$ is the spatial projection tensor $h^{ab} = g^{ab} +
u^{a}u^{b}/c^{2}$. The scalars $n,\,E,\,P$ can be, respectively,
identified with the particle number and matter--energy densities
and the equilibrium pressure. Inserting (\ref{f0_def}) into
(\ref{Liou_eq}) yields the constraint
  \begin{equation} p^a\,\alpha_{,a}+\beta_{(a;b)}\,p^a\,p^b \ = \
- m\,\left[\beta_a\,F^a + \frac{\partial F^a}{\partial
p^a}\right].\label{force_constr}  \end{equation}
In order to make further progress, we have to introduce
assumptions about $F^a$. Guided by the analogy to the Lorentz
force, we shall consider a force that is linear in the particle
momentum, i.e.
\begin{equation}
F^a (x,p) = A^{ab}(x)p_b  \ , \label{force ansatz}
\end{equation}
where $A^{ab}(x)$ is a function of the spacetime coordinates but
does not depend on the particle momentum $p^a$. Since the force
has to satisfy $p_aF^a = 0$, it follows that, analogously to the
Lorentz-force, $A_{ab} = - A_{ba}$ is valid. As a consequence we
have $\partial F^a/\partial p^a = 0$ and the condition
(\ref{force_constr}) decomposes into
  \begin{equation} \alpha_{,a} \ = \ -\ m\,\beta^c A _{ca}\ ,
  \qquad \beta_{(a;b)}
\equiv \frac{1}{2}{\cal L}_\beta g_{ab}\ = \ 0 \ . \label{Lie_1}
\end{equation}
As opposed to the case of the Lorentz force, the quantity
$A^{ab}(x)$ is not specified here. We only require that it should
be compatible with the integrability condition $\alpha_{,a;b} =
\alpha_{,b;a}$ which holds for spherically symmetric static
spacetimes to be discussed below. The second relation in
(\ref{Lie_1}) implies that $\beta^a \equiv u^a/k_{B}T$ is a
timelike Killing vector, which gives rise to the relations
\begin{equation}\Theta \equiv u^a_{;a} = \dot T = 0 \ ,
\label{stationarity}\end{equation}
where $T$ is the equilibrium temperature and
\begin{equation}\dot{u}_a + \frac{\nabla_a T}{T}= 0 \ ,
\label{Tolman}\end{equation}
where $\nabla_a T \equiv h_{a}^{b}T_{;b}$. Because of the
antisymmetry of $A^{ab}(x)$ the first of the relations
(\ref{Lie_1}) splits into
\begin{equation}\dot\alpha = 0 \ ,\qquad \nabla_a \alpha \ =
\frac{m}{k_{B}T} \, u^c \,h^m_a \,A_{cm}\ . \label{nabalpha}
\end{equation}
Using the force (\ref{force ansatz}) in the energy-momentum
balance  (\ref{Tbal_def}), we obtain
\begin{equation}
\left(E + P\right) \,\dot{u}_n + \nabla_n \,P = n\,m\,u^c \,h^m_n
\,A_{cm}\, = P \nabla_n\,\alpha \ . \label{mom bal}
\end{equation}
The projection of (\ref{Tbal_def}) in direction of $u^a$ is
satisfied identically,  since $\dot E=\Theta=0$ and the right hand
side of the resulting equation also vanishes.

It is interesting to consider  the Gibbs-Duhem relation
\begin{equation}
\mbox{d} P = \left(E+ P\right)\frac{\mbox{d} T}{T} + n T \mbox{d}
\left(\frac{\mu }{T} \right)\ ,  \label{Gibbs Duhem}
\end{equation}
where $\mu$ is the chemical potential. With the identification
(\ref{alpha_def}) this yields
\begin{equation}\nabla_n\,P = \left(E+P\right)\frac{\nabla_n T}{T}+
P\nabla_n\,\alpha \  , \label{Gibbs Duhem2}\end{equation}
which is completely general and quite independent of the action of
a force.  Obviously, the consistency between the last relation and
the momentum balance (\ref{mom bal}) is guaranteed through
(\ref{Tolman}).

The energy-momentum tensor $\Teq^{ab}$ is not conserved and
therefore it is not a suitable quantity in Einstein's field
equations. Following earlier considerations in a cosmological
context~\cite{gasQ1,gasQ2,gasQ3}), we try to map the ``source''
terms in the balances for $\Teq^{ab}$ on imperfect fluid degrees
of freedom of a conserved energy-momentum tensor of the form
\bse\ba T^{ab} &&\ = \ \Teq^{ab}+\tau^{ab}\ ,\label{full T}\\
\tau^{ab} &&\ = \ \Pi\,h^{ab} + \Sigma^{ab} +
2\,q^{(a}\,u^{b)}\ ,\label{full_tau}\\
\textrm{with}\quad u_a\,q^a &&\ = \ u_a\,\Sigma^{ab} \ = \ 0\
,\quad \Sigma^a\,_a \ = \ 0\ ,\label{qP}\ea\ese
where we can identify $\Pi$ as the scalar off-equilibrium
pressure, $\Sigma^{ab}$ as the traceless anisotropic stress tensor
and $q^a$ as the energy flux vector. Since we must have
$T^{ab}\,_{;b}=0$, then
  \begin{equation} \tau^{ab}\,_{;b} \ = \ - \Teq^{ab}\,_{;b}\ ,  \end{equation}
which in the present case without an energy flux reduces to
  \ba h^a\,_b\,\tau^{bc}\,_{;c} &=& \ \nabla^a\,\Pi  +
 \Pi\,\dot u^a  +
\dot u_b\,\Sigma^{ab}  +  h^b\,_c\,\Sigma^{ca}\,_{;b}\nonumber \\
&=& -P\ \nabla^a\,\alpha \ . \label{}   \ea
All these relations do not depend on a specific choice of the
antisymmetric quantity $A ^{ab}$ in (\ref{force ansatz}).
Consequently, on the fluid level the additional interaction
manifests itself through the appearance of the additional pressure
contributions $\Pi$ and $\Sigma^{ab}$.

In the following we are interested in the limit in which the
``equilibrium'' state variables introduced in (\ref{NTf0_def})
satisfy the equation of state of a non-relativistic
Maxwell-Boltzmann (MB) gas~\cite{kinetic} with rest mass--energy
density $\rho$,
\begin{equation} E \ = \rho\,\left(1+\frac{3}{2\,z}\right)\ ,\quad P \ = \
\frac{\rho}{z}\ ,\label{rhoP_def}\end{equation}
with
\begin{equation}
 n \ = \ \left[
\frac{mc}{\sqrt{2\pi\,z}\hbar}\right]^{3}\,\exp(\alpha -z)\ ,\quad
\rho \ = \ m\,n\,c^{2} ,\label{n_def} \end{equation}
where
  \begin{equation} z \ \equiv \ \frac{m\,c^{2}}{k_{B}T}\label{z_def}
  \end{equation}
is the ``relativistic coldness'' parameter. Therefore, $\alpha$ is
the fugacity scalar that is related to the chemical potential
$\mu$ by
  \begin{equation} \alpha \ = \ \frac{\mu}{k_{B}T} \ = \ \frac{\mu\,z}{mc^{2}}\
.\label{alpha_def}  \end{equation}

\section{Spherically symmetric and static spacetime}

We now apply the formalism outlined so far to a spherically
symmetric and static spacetime, whose metric can be given as
  \ba \dd s^2 && = \ -\exp\,(2\Phi/c^2)\,c^2\,\dd t^2 +
\left[1-\frac{2\,G\,M}{c^2\,r}\right]^{-1}\,\dd r^2\nonumber\\
&&+r^2\left(\dd
\theta^2+\sin^2\theta\,\dd
\phi^2\right)\ ,\label{ss_metric}  \ea
where $\Phi=\Phi(r)$ and $M=M(r)$ have, respectively, units of
velocity squared and mass. Our strategy is to combine the
corresponding field equations with the equilibrium conditions of
the previous section. Then the 4--velocity and the 4--acceleration
(in velocity and acceleration units) take their customary forms
\ba u^a \ = \ \exp(-\Phi/c^2)\,c\,\delta^a_0,\qquad \dot u_a \ = \
\Phi\,'\,\delta^r_a\ .\label{uudot}\ea
Other relevant quantities are
\bse
\ba
%z  && = \ \frac{m\,c^2}{k_{_{B}}\,T}\ ,\\
\beta^a &&  = \
 \frac{c\exp(-\Phi/c^2)}{k_{_{B}}\,T}\,\delta^a_0\ , \\
\frac{z_{;a}}{z}
&& = \ \frac{z'}{z}\,\delta^r_a \ = \ -\frac{T'}{T}\,\delta^r_a \
,\label{objects}  \ea\ese
where a prime denotes derivative with respect to $r$. Equation
(\ref{Tolman}) implies the well known Tolman law $T\propto
\sqrt{-g^{00}}$, which remains valid here for any choice of
$A^{ab}$ in (\ref{force ansatz}).

Particle motion in a spherically symmetric field is characterized
by conservation of the of the angular momentum
$L_0^2=r^4[(p^\theta)^2+\sin^2\theta\,(p^\phi)^2] $. But with $F^a
\neq 0$ the particles do not follow geodesics and so
$\sqrt{-g_{00}}\,p^0 $ is not a constant of motion. In the present
case the force may be written as  $F^a = F^0\,\delta^a_0+
F^r\,\delta^a_r$, hence the two nonzero components  of
$F^a=A^{ab}\,p_b/c = (A/c)\, p_{[0}\delta_{r]}^a$ are
\bse\label{force comp}\ba F^0  &=&  -A\,\frac{p_r}{c}  =
-m\,A\,g_{rr}\,\frac{v^r}{c},\\
F^r  &=&  A\,\frac{p_0}{c}  \nonumber\\
&=&
-m\,A\,\textrm{e}\,^{\Phi/c^2}\,\left[1+g_{rr}\frac{(v^r)^2}{c^2}
+\frac{L_0^2}{m^2\,c^2\,r^2}\right]^{1/2} \ ,\nonumber\\\ea\ese
where
\begin{equation}A \ = \ A(r) \ \equiv \ A^{r0} \ = \
-A^{0r}\label{Fdef}\end{equation}
and we have eliminated $p_0$ from $g_{ab}p^ap^b=-m^2c^2$ and used
$p^r=m\,v^r $, as well as conservation of angular momentum. The
equilibrium condition (\ref{nabalpha}) for $\alpha$ becomes
  \begin{equation} \alpha\,' \ = \ \frac{\textrm{e}^{\Phi/c^2}\,\,r}{r -
2 G M/c^2}\,\frac{m\,A}{\kB\,T} \ .\label{alphaprime0}
\end{equation}
Notice that $A$ and $F^a$ have, respectively, units of
acceleration and force.

\section{Field equations}

For the metric (\ref{ss_metric}), we have $q^a=0$ and the most
general form of the spacelike traceless tensor is given by
$\Sigma^0\,_0=0,\,\, \Sigma^r\,_r=-2\,\Sigma=
\,-2\,\Sigma^\theta\,_\theta = -2\,\Sigma^\phi\,_\phi$,\,  where
$\Sigma=\Sigma(r)$ is an arbitrary function. Thus, the
momentum--energy tensor is
\ba T^0\,_0 && = \ -E,\nonumber\\  T^r\,_r && = \ P+\Pi-2\,\Sigma\ ,\nonumber\\
T^\theta\,_\theta \ = \ T^\phi\,_\phi && = \
P+\Pi+\Sigma,\nonumber\\
\label{Tab_ss}\ea
leading for (\ref{ss_metric}) and (\ref{Tab_ss}) to the field
equations (or ``equilibrium'' TOV type equations)
\ba M'\ && = \ 4\,\pi\,(E/c^2)\,r^2,\label{eqMr}\\
\Phi\,'\ && = \
G\,\frac{M+\,(P+\Pi-2\,\Sigma)\,r^3/c^2}{r\,(r-2\,G\,M/c^2)}\ ,
\label{eqPhr}\ea
where, from (\ref{rhoP_def}) and (\ref{n_def}), we know that $E$
and $P$ are functions of $z$ and $\alpha$. This has to be
complemented by (\ref{alphaprime0}) and by
  \begin{equation} \frac{z'}{z} \ = \ \frac{\Phi\,'}{ c^2}\ ,\label{eqzr}
\end{equation} %% which follows from (\ref{Tolman}), as well as by
  \begin{equation} (\Pi-2\,\Sigma)\,'
 = \
-\alpha\,'\,P- (\Pi-2\,\Sigma)\,\,
\frac{\Phi\,'}{c^2}-\frac{6\,\Sigma}{r}\ .\label{eqPpr}  \end{equation}
In the absence of the additional interaction, i.e. for
$\Pi=\Sigma=0$, we recover the standard TOV equations while
Eq.~(\ref{eqPpr}) reduces to $\alpha\,' = 0$.

 \noindent Assuming $A$ to be known, equations
(\ref{eqMr})--(\ref{eqPpr}), complemented by (\ref{rhoP_def}) and
(\ref{n_def}), constitute a system of five differential equations
for six unknowns: $M,\,\Phi,\,z,\,\alpha,\,\Pi,\,\Sigma$. Hence,
an extra constraint is needed to render this system determinate.
One way of achieving this determinacy is to provide a specific
relation between $\Pi$ and $\Sigma$. The simplest cases are either
one of
\bse\ba \Sigma && = \ 0,\qquad \textrm{purely isotropic case}\ ,\\
\Pi && = \ 0, \qquad \textrm{purely anisotropic case}\ea\ese
Another possible approach follows by defining ``radial'' and
``tangential'' pressures as
\begin{equation} P_r \ = \ \Pi-2\,\Sigma,\qquad P_\perp \ = \
\Pi+\Sigma,\label{P_ss}\end{equation}
leading to a system like (\ref{eqMr})--(\ref{eqPpr}), with (\ref{eqPhr}) and
(\ref{eqPpr}) replaced by
\ba\Phi\,' &&  = \
G\,\frac{M+\,(P+P_r)\,r^3/c^2}{r\,(r-2\,G\,M/c^2)}\
, \label{eqPhrss}\\
P_r\,'
 && = \
-\alpha\,'\,P- P_r\,\,\left(
\frac{\Phi\,'}{c^2}+\frac{2\,\Gamma}{r}\,\right)\ .\label{eqPrss}\ea
where
  \begin{equation} \Gamma \ = \ 1 - \frac{P_\perp}{P_r},  \end{equation}
is the ``anisotropy factor''. If the latter is selected (say, from empirical
considerations) we have also a determined system.

\section{Linear isotropic case}

We consider now the  case in which $F^a$ is linear in the momentum $p^a$,
with purely isotropic stresses characterized by $\Sigma=0$
(or $P_r=P_\perp $). The system
(\ref{eqMr})--(\ref{eqPpr}) reduces to
\bse\ba M'\ && = \ 4\,\pi\,\rho\,\left(1+\frac{3}{2\,z}\right)\,r^2\
,\label{eqMr2}\\
\Phi\,' \ && = \
G\,\frac{M+\,(\rho\,z^{-1}+\Pi/c^2)\,r^3}{r\,(r-2\,G\,M/c^2)}\ ,
\label{eqPhr2}\\
\alpha\,' \ && = \   \frac{z\,r\, \hbox{e}^{\Phi/c^2}}{r - 2\,G\,
M/c^2}\,\frac{A}{c^2} \ ,
\label{eqar2}\\
\Pi\,' \ &&= \
-\alpha\,'\,\frac{\rho\,c^2}{z}-\Pi\,\frac{\Phi\,'}{c^2}\
,\label{eqpir2}\ea\ese
where we used (\ref{rhoP_def}) and (\ref{n_def}) to eliminate $E$ and $P$
in terms of $z$ and $\rho$, with $\rho=\rho(\alpha,z)$   and $z=z(\Phi)$
given by
\bse\ba \rho \ &&= \
\rho_c\,\left(\frac{z_c}{z}\right)^{3/2}\,\exp\,(\alpha -
\alpha_c+ z_c-z)\ ,\label{eqrho2}\\ z \ &&= \
z_c\,\exp\,\left[(\Phi-\Phi_c)/c^2\right]\ ,\label{eqz2}\ea\ese
where the subscript $_c$ will denote henceforth evaluation at the symmetry
center $r=0$. It is convenient for self gravitating gas sources to eliminate
$z_c$ in terms of the velocity dispersion of gas particles~\cite{BT}
\ba \sigma_c^2 \ = \ \frac{c^2}{z_c} \ = \ \frac{\kB\,T_c}{m}\ ,
\label{eqsigmac}\ea
and to express $\Phi$ in terms of the dimensionless variable
\ba \Psi \ \equiv \ \frac{\Phi_c-\Phi}{\sigma_c^2}\
,\label{Psi_def}\ea
which reduces to the ``normalized'' Newtonian potential in the
Newtonian limit. Equations (\ref{eqrho2}) and (\ref{eqz2}) then
become
\bse\ba Y \ \equiv \ \frac{\rho}{\rho_c} &=&
\exp\,\left[\,\alpha-\alpha_c+\frac{3}{2}\xi\,\Psi
+\frac{1}{\xi}\left(1-\hbox{e}^{-\xi\,\Psi}\right)\right],\nonumber\\
\label{eqrho3}\\ z &=&
\frac{1}{\xi}\,\hbox{e}^{-\xi\,\Psi},\label{eqz3}\ea\ese
where
\begin{equation} \xi \ \equiv \ \frac{\sigma_c^2}{c^2}\ .\label{xi_def}
\end{equation}
We introduce  the dimensionless variables
\bse\label{MPA_def}\ba x \ &\equiv & \ \frac{r}{r_0}\ ,\\ r_0^{-2} \ &=& \
\frac{4\pi\,G\,\rho_c}{\sigma_c^2} \ = \
\frac{3}{2}\,\frac{h^2H_0^2}{\sigma_c^2}\,\frac{\rho_c}{\rhocr}\
,\label{x_def}\\
\M \ &\equiv & \ \frac{M}{4\pi\,\rho_c\,r_0^3}\ ,\\
\CP \ & \equiv & \ \frac{\Pi}{\Pi_c}\ ,\qquad
 \nu \  \equiv  \ \frac{\Pi_c}{\rho_c\,c^2} \label{nu_def}\\
\A & \equiv & \ \
\frac{A\,r_0\,\hbox{e}^{\Phi_c/c^2}}{\sigma_c^2}\ = \
\frac{2}{3}\,\frac{\rhocr}{\rho_c}\,\frac{A\,\textrm{e}^{\Phi_c/c^2}}{r_0\,h^2
H_0^2}\ ,\nonumber\\
\ea\ese
where we have used:
$$\rhocr=3h^2H_0^2/(8\pi G),\quad\xi=(4\pi
G\rho_c/c^2)\,r_0^2\ .$$
With the definition (\ref{eqsigmac}) the parameter $\xi$ in
Eq.~(\ref{xi_def}) represents a dimensionless velocity dispersion
of the particles. A corresponding quantity, the parameter $\nu$
defined in Eq.~(\ref{n_def}), arises due to the existence of the
additional pressure $\Pi$. Both these parameters will be relevant
for the Newtonian type limits discussed below. In terms of the
parameters defined in (\ref{MPA_def}) the system
(\ref{eqMr2})--(\ref{eqpir2}) becomes
\bse\label{eqsrel}\ba \frac{\dd\,\M}{\dd\,x} \ &&= \
Y\,\left[1+\frac{3}{2}\,\xi\,\,\hbox{e}^{\,\xi\,\Psi}\right]
\,x^2\ ,\label{eqMx}\\
\frac{\dd\,\Psi}{\dd\,x} \ &&= \
-\frac{\M+\left[\,\xi\,Y\,\textrm{e}\,^{\xi\,\Psi}+\nu\,
\CP\right]\,x^3}{x\,(x-2\,\xi\,\M)}\ ,
\label{eqPsix}\\
\frac{\dd\,\alpha}{\dd\,x} \ &&= \
\A\,\frac{\hbox{e}^{-\,2\,\xi\,\Psi}\,x}{x-2\,\xi\,\M}
\ ,\label{eqalphax}\\
\frac{\dd\,\CP}{\dd\,x} \ &&= \ \xi\,\CP\,\frac{\dd\,\Psi}{\dd\,x}
-\,\,\frac{\xi}{\nu}\,Y\,\hbox{e}^{\,\xi\,\Psi}\,\frac{\dd\,\alpha}{\dd\,x}
\ ,\label{eqPix}\ea\ese
where we have eliminated $z$ from (\ref{eqz3}), while $Y$ follows
from (\ref{eqrho3}). We emphasize again, that this set of equation
is valid for any force of the type (\ref{force ansatz}). Although
for an explicit evaluation we might propose any functional form
$\A=\A(r)$, it is more convenient to provide a covariant form for
$A^{ab}$.

\section{Specific forces}

While the assumed structure (\ref{force ansatz}) of the force
$F^a$ shares some features of the Lorentz force, its explicit
functional form has been left open. The accelerations $A^{ab}$ are
only restricted by the symmetry requirement  $A^{ab} = - A^{ba}$.
Since we expect this force to model an effective interaction, it
may self-consistently depend on the macroscopic fluid quantities
in addition to its (fixed, linear) dependence on the microscopic
particle momentum.

\subsection{First case}

A simple choice for an antisymetric tensor $A^{ab}$ for a spherically symmetric
fluid spacetime is
\begin{equation}
A^{ab} = \frac{2 \,\lambda}{c} \,\dot u ^{[a}\,u^{b\,]}\ , \label{Fab1}
\end{equation}
which, for the metric (\ref{ss_metric})
and using the variables (\ref{MPA_def}), yields
\ba
\A &=& \, -\lambda
\,\textrm{e}\,^{\xi\,\Psi} \, \frac{\dd\,\Psi}{\dd\,x}\left(1 -
\frac{2\,\xi\M}{x}\right)\nonumber\\
&=&
\lambda\,\,\textrm{e}\,^{\xi\Psi}\,\left[\frac{\M}{x^2}
+\left(\,\xi\,Y\,\textrm{e}\,^{\xi\,\Psi}+\nu\,
\CP\right)\,x\right],
\label{Frt}
\ea
while the components of $F^a$ are
\bse\ba F^0 \ &=& \
-\lambda\,m\,\Phi'\,g^{rr}\,\textrm{e}^{-\Phi/c^2}\,\frac{v^r}{c}\ ,\\
F^r \ &=& \
-\lambda\,m\,\Phi'\,g^{rr}\,\left[1+g_{rr}\frac{(v^r)^2}{c^2}
+\frac{L_0^2}{m^2\,c^2\,r^2}\right]^{1/2}.\nonumber\\
\ea\ese
Thus, the dimensionless quantity $\lambda$ is the free parameter
(not necessarily constant) which determines the functional form of
the force strength.

Consequently, the set of relativistic equations to be solved is
(\ref{eqsrel}), with $\A$ given by (\ref{Frt}) so that
(\ref{eqalphax}) and (\ref{eqPix}) are replaced by
\bse\ba
\frac{\dd\,\alpha}{\dd\,x} \ &=& \ -\lambda \,
\hbox{e}^{-\,\xi\,\Psi}\,\frac{\dd\,\Psi}{\dd\,x} \
,\label{eqalphax1}
\\
\frac{\dd\,\CP}{\dd\,x} \ &=& \
\left[\,\xi\,\CP+\frac{\xi}{\nu}\,\lambda\,Y\,\right]\,
\frac{\dd\,\Psi}{\dd\,x}\ .\label{eqPix1}\ea\ese

\subsection{Second case}

An interaction mediated by the curvature provides another choice
of an antisymmetric tensor $A^{ab}$:
\begin{equation}
A^{a b} = \frac{\ell\,^2}{c} \,\,g^{b c}\,R^a_{\ c d e}\,\dot u ^{d}
\,u^{e}\ , \label{Fab2}
\end{equation}
where $\ell$ is a characteristic length scale and  $R^a_{\ cde}$
are the components of the curvature tensor, hence the
corresponding interaction is non--minimal. Now, forces which are
proportional to the curvature tensor are usually not admitted in
General Relativity since they violate the equivalence principle.
Here, it is the self-consistent mapping of the corresponding
source terms on imperfect fluid degrees of freedom of a conserved
energy momentum tensor according to (\ref{full T}) - (\ref{qP}),
which allows us to consider this force as an internal interaction
within Einstein's theory.

The choice (\ref{Fab2}) leads to
\begin{eqnarray}
A &=&  \ell^2 \,g^{tt}\,R^r_{\ trt}\,\dot u ^{r}\,u^{t}/c\
\nonumber\\
&=& \ell^2\, g^{rr}\,\Phi'\,\textrm{e}^{-\,\Phi/c^2}
\left[g^{rr}\left(\frac{\Phi''}{c^2} +
\frac{\Phi'{}^2}{c^4}\right) +
\frac{(g^{rr})'}{2}\,\frac{\Phi'}{c^2}\right] ,\nonumber\\
\label{Frt2}
\end{eqnarray}
so that the force components are $F^0=-A\,p_r/c$ and
$F^r=A\,p_0/c$. With the help of (\ref{MPA_def}) we obtain the
dimensionless quantity
\ba\A \ = \
\CS\,\textrm{e}\,^{\xi\Psi}\left[\,\frac{\M}{x^2}+\left(Y\xi\textrm{e}\,
^{\xi\Psi}+\nu\CP\right) x\,\right]\,\times\nonumber\\
\left\{\frac{2\M}{x^3}-\left[Y\left(1+\frac{5}{2}\xi\textrm{e}\,^{\xi\Psi}\right)
+\nu\CP\right]\right\}\ ,\label{Frt3}\ea
where
\begin{equation}\CS \ = \ \frac{\ell^2\,\xi}{r_0^2} \ = \ \frac{4\pi G
\rho_c}{c^2}\,\ell\,^2 \ = \
\frac{2}{3}\,\frac{h^2\,H_0^2\,\ell^2}{c^2}\,\frac{\rho_c}{\rhocr}\
.\label{Sdef}
\end{equation}
The relativistic equilibrium equations to be solved are then
(\ref{eqsrel}) with $\A$ given by (\ref{Frt3}), hence
(\ref{eqalphax}) and (\ref{eqPix}) must be replaced by
\bse\ba
\frac{\dd\,\alpha}{\dd\,x} \ &=& \ -\CS \,
\hbox{e}^{-\,\xi\,\Psi}\,\Delta\,\frac{\dd\,\Psi}{\dd\,x} \
,\label{eqalphax2}
\\
\frac{\dd\,\CP}{\dd\,x} \ &=& \
\left[\,\xi\,\CP+\frac{\xi\,\CS}{\nu}\,Y\,\Delta\right]\,
\frac{\dd\,\Psi}{\dd\,x}\ ,\label{eqPix2}\ea\ese
where
\begin{equation}\Delta \ \equiv \ \frac{2\M}{x^3}-
\left[Y\left(1+\frac{5}{2}\xi\textrm{e}\,^{\xi\Psi}\right)
+\nu\CP\right]\ .\end{equation}
 \\

\section{Newtonian type limits}

Since our approach implies a non-standard degree of freedom, we
have to define what we mean by a Newtonian type limit. The free
parameter $\xi$ defined in (\ref{xi_def}) is associated with the
characteristic velocities of a spherically symmetric MB gas. The
second free parameter $\nu$ in (\ref{MPA_def}) is new and provides
the velocity dispersion $\Pi/\rho$ associated with the additional
pressure $\Pi$. It quantifies the impact of the non-standard
interaction on our equilibrium configuration. We expect this
impact to be a small correction to the standard case (the case
without additional interaction), hence $\nu \ll 1$. Since the
dimensionless quantities $\Psi,\,\M,\,\alpha$ and $\CP$ are not
necessarily ``small'', it is reasonable to consider a weak field
limit based on the conditions
  \begin{equation}\frac{\Phi_{c}}{c^{2}}\ \ll \ 1\ ,\qquad\xi \ \ll \ 1
  \qquad\underline{\textrm{and}}\qquad  \nu \ \ll \
\ 1\ ,\label{newtcond}
\end{equation}
irrespective of how $\xi$ and $\nu$ are related. The metric
functions in (\ref{ss_metric}) become up to first order in $\xi$
\ba \sqrt{-g_{00}} \ = \ \hbox{e}^{\Phi_c/c^2-\xi\,\Psi} \ \approx
\
\hbox{e}^{\Phi_c/c^2}\,\left[1-\xi\,\Psi+O(\xi^2)\right],\nonumber\\
\sqrt{g_{rr}} \ = \ \left[1-\frac{2\,\xi\,\M}{x}\right]^{-1/2} \
\approx \
1+\,\xi\,\frac{\M}{x}+O(\xi^2)\ ,\nonumber\\
\ea
while applying (\ref{newtcond}) to (\ref{eqrho3}) we obtain
  \begin{equation} Y \ = \ \frac{\rho}{\rho_c} \ \approx \
\hbox{e}^{\alpha-\alpha_c+\Psi}\,+O(\xi).\label{eqrho4}  \end{equation}
Newtonian type limits imply Newtonian particle velocities $v^r\ll
c$ and $L_0 \ll mc$. Under the conditions (\ref{newtcond}) we also
have $g_{rr}\approx 1$ and $g_{00}\approx -1$, hence the
components of $F^a$ given by (\ref{force comp}) become
\begin{equation}F^r \approx -m\,A\ ,\qquad F^0\approx
F^r\,\frac{v^r}{c}\ ,\label{force comp newt}\end{equation}
so that $F^0\ll F^r$. Conditions (\ref{newtcond}) imply a weak
field limit which can be  identified with various types of
``nearly Newtonian'' conditions that are  obtained by comparing
$\xi$ and $\nu$ and by expanding all incumbent variables with
respect to either one (or both) of these dimensionless ratios.

\subsection{The ``standard'' Newtonian limit}

In the general relativistic treatment of the ``standard'' MB gas
(without self--interaction mediated by $F^a$, so that:
$\alpha-\alpha_c=\nu=0$) the Newtonian limit follows by expanding
all relevant quantities up to first order in $\xi$. This leads to
the ``isothermal sphere'' with relativistic corrections of order
$\xi$. An equivalent Newtonian limit can be defined for the case
$F^a\ne 0$ if together with (\ref{newtcond}) we have
\begin{equation} \frac{\xi}{\nu} \ = \ k, \qquad k\ > \ 1, \qquad
\textrm{but}\quad k \ \sim
\ O(1),
\label{nullxi}\end{equation}
so that $\rho_c\,\sigma_c^2=k\,\Pi_c$ and the relativistic
correction introduced by $\Pi$ is of the same order but smaller
than the thermal one, corresponding to the equilibrium hydrostatic
pressure $P$. By expanding up to $O(\xi)$ we get modifications of
the isothermal sphere characterized by the post--newtonian
equilibrium equations
\bse\label{eqs0x}\ba \frac{\dd\,\M}{\dd\,x} \ &=& \
Y\,x^2\, + O(\xi) ,\label{eqM0x}\\
\frac{\dd\,\Psi}{\dd\,x} \ &=& \
-\frac{\M}{x^2} + O(\xi),
\label{eqPsi0x}\\
\frac{\dd\,\alpha}{\dd\,x} \ &=& \ \A + O(\xi),\label{eqalpha0x}\\
\frac{\dd\,\CP}{\dd\,x} \ &=& \
-\,\,k\,\,Y\,\,\A  + O(\xi)
 ,\label{eqPi0x}\ea\ese
with $Y$ given by (\ref{eqrho4}).

\subsubsection{First case}

If we assume (\ref{Fab1}), we have in the weak field limit
\begin{equation}\A= \lambda\,\frac{\M}{x^2}+O(\xi)\end{equation}
and so the components of the force $F^a$ are
\begin{equation} F^r\approx
-\lambda\,\frac{GMm}{r^2},\qquad F^0 \approx F^r\,\frac{v^r}{c}\
,\label{FNewton}
\end{equation}
where we have eliminated $\Phi'$ from (\ref{eqPhr2}). The
Newtonian structure of the force in this limit is obvious. The
equilibrium equations are then (\ref{eqs0x}), with
(\ref{eqalpha0x}) and (\ref{eqPi0x}) replaced by
\bse\ba\frac{\dd\,\alpha}{\dd\,x} \ &=& \
\lambda\,\frac{\M}{x^2}+O(\xi)\ ,\label{alphax_WL1}\\
\frac{\dd\,\CP}{\dd\,x} \ &=& \
-k\,\lambda\,Y\,\frac{\M}{x^2}+O(\xi)\ .\label{Pi_WL1}\ea\ese
The fact that the additional interaction, introduced in a general
relativistic setting, reduces to an additional Newtonian type
force (\ref{FNewton}) in the appropriate limit reminds of
corresponding features of Modified Newtonian Dynamics
(MOND)~\cite{MOND}. While the latter, however, was introduced as
an alternative to DM, the force describes a self-interaction
within the DM in our case. Nevertheless, our result does not seem
to exclude an approach in which the DM is replaced by baryonic
matter with a modified Newtonian limit of a general relativistic
description, where the modification is a consequence of a
non-geodesic motion of the (baryonic) particles of the medium. We
plan to follow this line in future work.

\subsubsection{Second case}

For the curvature mediated force we have
\begin{equation}\A =
\frac{\CS\,\M}{x^3}\,\left(\frac{2\,\M}{x^3}-Y\right)+O(\xi)\ ,
\end{equation}
hence, the components of $F^a$ are
\begin{equation}F^r\approx
-\frac{\ell\,^2}{c^2}\,\frac{GMm}{r^2}\left[\frac{2GM}{r^3}-4\pi\,
G\,\rho\right],\quad F^0\approx F^r\,\frac{v^r}{c}\
.\end{equation}
\\
The equilibrium equations are (\ref{eqs0x}), but now with (\ref{eqalpha0x})
and (\ref{eqPi0x}) replaced by
\bse\label{WL2}\ba\frac{\dd\,\alpha}{\dd\,x} \ &=& \
\CS\,\left[\frac{2\,\M}{x^3}-Y\right]\frac{\M}{x^2}+O(\xi)\ ,
\label{alphax_WL2}\\
\frac{\dd\,\CP}{\dd\,x} \ &=& \
-k\,\CS\,Y\,\left[\frac{2\,\M}{x^3}-Y\right]\,\frac{\M}{x^2}
+O(\xi)\ .\label{Pi_WL2}\ea\ese

\subsection{Newtonian limit with a repulsive force}

If instead of (\ref{nullxi}) we choose (\ref{newtcond}) together with
\begin{equation}\nu \ < \ 0\ ,\qquad |\nu| \ = \
b\,\xi \ > \ \xi\ ,\label{LambdaConds}\end{equation}
then, instead of (\ref{eqs0x}), the equilibrium equations in the
Newtonian limit are
\bse\label{eqs1x}\ba \frac{\dd\,\M}{\dd\,x} \ &=& \
Y\,x^2\, + O(\xi)\ ,\label{eqM1x}\\
\frac{\dd\,\Psi}{\dd\,x} \ &=& \ -\frac{\M}{x^2}
+|\nu|\,\CP\,x+O(\xi)\ ,
\label{eqPsi1x}\\
\frac{\dd\,\alpha}{\dd\,x} \ &=& \ \A + O(\xi)\ ,\label{eqalpha1x}\\
\frac{\dd\,\CP}{\dd\,x} \ &=& \
\,\,\xi\,\CP\,\,\frac{\dd\,\Psi}{\dd\,x} +\,\frac{1}{b}\,\,Y\,\,\A
+ O(\xi^2) \ ,\label{eqPi1x}\ea\ese
with $Y$ given by (\ref{eqrho4}). If $1/b\ll\xi$, then equation (\ref{eqPi1x})
implies $\CP \propto \,\textrm{e}^{\xi\Psi}=1+\xi\Psi+O(\xi^2)$, hence
\begin{equation}\Pi \ = \ \Pi_c\,\CP \ = \ \Pi_c\,\left[\,
1 +O(\xi)\,\right] \ \approx \ \Pi_c\ .\end{equation}
But if $1/b\gg\xi$ (a more likely outcome), then the form of $\CP$ cannot be
guessed before a numerical integration. Still, the term $|\nu|\,\CP\,x$ in
(\ref{eqPsi1x}) behaves as a repulsive force similar to a positive ``cosmological
constant''  that can be associated with the matter--energy
density~\cite{SussHdez}
\begin{equation}|\Pi_c| \approx
h^2\,\Omega_\Lambda\,\rhocr\,c^2\ .\label{LambdaE}\end{equation}
The equilibrium equations in each case follow by inserting in
(\ref{eqs1x}) the appropriate form of $\A$ and taking $\nu<0$. For
the first case we have
\begin{equation}\A \ = \ \lambda\,\left[\frac{\M}{x^2}-|\nu|\,\CP\,
x\right],\end{equation}
with
\begin{equation} F^r \approx
-\lambda\,G\,m\,\left[\frac{M}{r^2}-\frac{|\Pi|}{c^2}\,r\right]\,
.
\end{equation}
For the second case we have
\begin{equation}\A \ = \ \CS\,\left[\frac{\M}{x^2}-|\nu|\,\CP\,
x\right]\,\left[\frac{2\M}{x^3}-Y+|\nu|\CP\right],\label{Acase2}\end{equation}
with
\begin{equation}F^r \approx
-\frac{m\,\ell\,G^2}{c^2}\,\,\left[\frac{M}{r^2}
-\frac{|\Pi|}{c^2}\,r\right]\left[\frac{2M}{r^3}-4\pi\rho
+\frac{|\Pi|}{c^2}\right]\,.\end{equation}

\section{Numerical examples}

We examine separately two configurations, one for each form of the
generalized forces (\ref{Fab1}) and (\ref{Fab2}). These examples
aim at illustrating how the formalism we have introduced works in
practice. Although these examples are not meant to describe
``realistic'' models of equilibrium systems, they show how
generalized forces in collisionless gases might influence known
velocity and density profiles and hence are of
interest in astrophysical applications. \\

\subsection{A modified isothermal sphere}

Let us consider a generalized force complying with (\ref{Fab1}),
with the dimensionless parameter $\lambda$ given by
\begin{equation}\lambda \ = \ \frac{1}{1+x^2}\ .\end{equation}
In the ``standard'' Newtonian limit (\ref{newtcond}) and
(\ref{nullxi}) with $k=1$, we obtain hydrostatic configurations
that resemble the isothermal sphere (case $\alpha=\CP=0$ without
force). Assuming the parameters characteristic of a galactic dark
matter halo similar to that of the Milky Way~\cite{ranges}:
\begin{equation}\rho_c=0.01\,M_\odot/\textrm{pc}^3,
\qquad
\sigma_c=140\,\,\textrm{km/sec},\label{MWvals}\end{equation}
we integrate the equilibrium equations (\ref{eqM0x}),
(\ref{eqPsi0x}), (\ref{alphax_WL1}) and (\ref{Pi_WL1}) and show
the results in figure 1, where solid curves (marked as ``MIS'')
denote the modified isothermal sphere in comparison with the
curves corresponding to the isothermal sphere (dotted curves
marked as ``IS''). Figure 1(a) displays the rotation velocity
profile obtained from (\ref{eqPhrss}), (\ref{eqPsix}) and
(\ref{eqPsi0x})
\begin{equation}\frac{\Vrot^2}{\sigma_c^2} \ = \
\frac{r\,\Phi'}{\sigma_c} \ = \ -x\,\frac{\dd \Psi}{\dd x} \ = \
\frac{\M}{x}+O(\xi).\label{vrot1}\end{equation}
The profile $\Vrot$ is qualitatively analogous to the isothermal
one, but reveals slight differences from the latter already in the
region occupied by visible  matter (up to a radius of $\sim 30$
kpc): it is steeper near the center and reaches a higher maximal
velocity. Instead of the ``flat'' isothermal profile, the present
case shows a slight decay in $\Vrot$. Figure 1(b) shows the
density profile, $Y=\rho/\rho_c$, which reveals a wider ``flat
core'' region with the same $r^{-2}$ asymptotic decay of the
isothermal sphere. Figure 1(c) displays the function $\alpha$,
while figure 1(d) compares the curve for the pressure associated
with the generalized force, $\CP =\Pi/\Pi_c$ with that for the
hydrostatic pressure $P/P_c= Y\,\textrm{e}\,^{\xi\,\Psi}$. While
$P\propto Y$, as in the isothermal case, $\CP$ decays
exponentially.

It is worthwhile mentioning that similar models of modified
isothermal spheres can be obtained with the curvature coupled
force (\ref{Fab2}), though we will use this type of force to model
an effective cosmological constant.

\subsection{Cosmological constant from curvature coupling}

Assuming a curvature mediated force (\ref{Fab2}), we examine now
the case of a weak field limit that resembles an isothermal sphere
in equilibrium with a cosmological constant (\textit{i.e.} a
cosmological field with matter--energy density
$E_\Lambda=-P_\Lambda =\Omega_\lambda\,h^2\,\rhocr\,c^2$
\cite{SussHdez}).  This suggests a generalized force of cosmic
origin, characterized by very large length scales. In fact, from
(\ref{Sdef}) with $H_0=100$\,km/(sec\,Mpc),
\begin{equation}\CS \ = \ 7.4\,\times 10^{-8}\,
\frac{h^2\,\ell^2}{(\textrm{Mpc})^2}\,\frac{\rho_c}{\rhocr} \ = \
0.0015\,\frac{\ell^2}{(\textrm{Mpc})^2},\end{equation}
%
%\vfill
%\eject
%\begin{widetext}
\begin{figure}
\centering
\includegraphics[height=22cm]{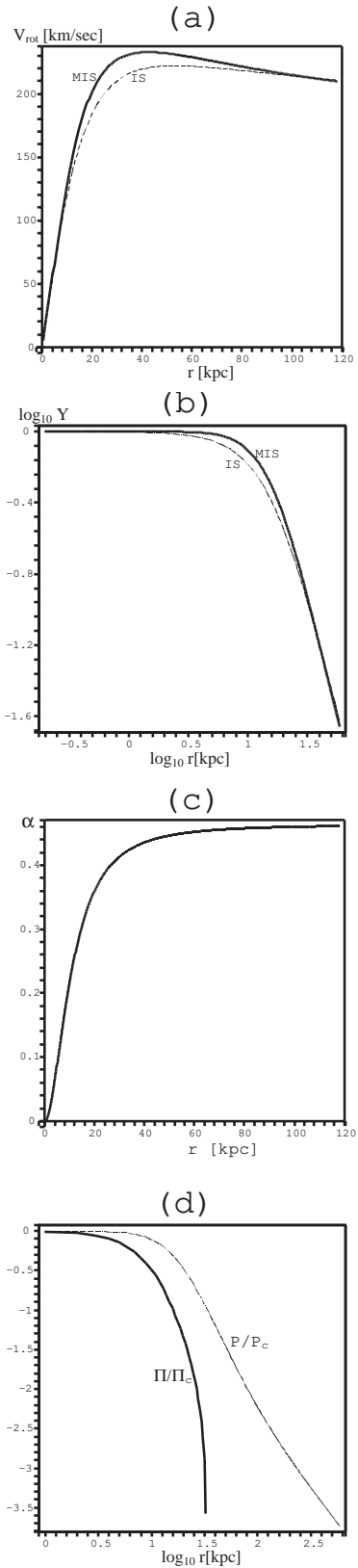}
%\picplace{5cm}{2cm} % Give the correct figure height and width in cm
%
\caption{Panels (a) and (b) depict the rotation velocity and
density profiles for the modified isothermal sphere (MIS), in
comparison with those of the isothermal sphere (IS), for a  Milky
Way size galactic halo under the assumption of the coupling
(\ref{Fab1}) with $\lambda=1/(1+x^2)$. Panel (c) displays the
function $\alpha(x)$ and panel (d) provides a comparison between
the hydrostatic pressure $P/P_c=Y\,\textrm{e}^{\xi\,\Psi}$ and the
pressure $\CP=\Pi/\Pi_c$ associated with the generalized force.
See section IX. A. for more details.}
\label{fig1}       % Give a unique label
\end{figure}
%\end{widetext}
%
%\begin{widetext}
\begin{figure}
\centering
\includegraphics[height=22cm]{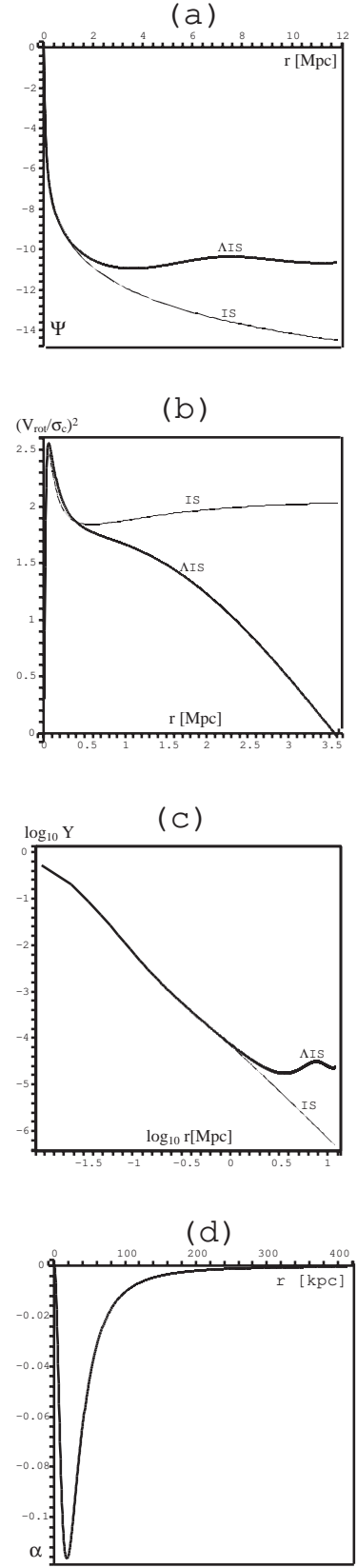}
%\picplace{5cm}{2cm} % Give the correct figure height and width in cm
%
\caption{Panels (a), (b) and (c) depict the curves (marked as
``$\Lambda$IS'') of the normalized potential $\Psi$, squared
rotation velocity and density profile for the case in which the
curvature coupled generalized force (\ref{Fab2}) acts as an
effective cosmological constant. The corresponding curves of the
isothermal sphere (IS) are shown for comparison. Panel (d)
displays the function $\alpha(x)$. Notice that $\Vrot^2=0$ at the
same radius ($r\approx 3.6$\,Mpc) where $\dd\Psi/\dd x=0$. See
section IX. B.  for more details.}
\label{fig2}       % Give a unique label
\end{figure}
%\end{widetext}
%
where we have used $h=0.7$ and (\ref{MWvals}), it follows
that $\CS=1$ corresponds to a characteristic length scale of $25.87$\,Mpc,
roughly an order of magnitude larger than the size of the largest
galactic clusters in virialized equilibrium (it is roughly the size of a
supercluster \cite{padma}). Following (\ref{LambdaE}), the ratio of
central hydrostatic pressure to the pressure of the $\Lambda$--field is
\begin{equation}b \ = \ \frac{|\nu|}{\xi} \ = \
\frac{h^2\,\Omega_\Lambda\,\rhocr\,c^2}{\rho_c\,\sigma_c^2} \ \approx \
40\end{equation}
where we have used (\ref{MWvals}), as well as $h=0.7$ and $\Omega_\Lambda=0.7$.
Hence, in order to have a curvature mediated force acting as a cosmological
constant, it is reasonable to assume $\CS=1$, together with conditions
(\ref{LambdaConds}) with $b=40$. Notice that the rotation velocity profile is no
longer given by (\ref{vrot1}), but by
\begin{equation}\frac{\Vrot^2}{\sigma_c^2} \ = \ \frac{\M}{x} -
|\nu|\,\CP\,x^2+O(\xi).\label{vrot2}\end{equation}
We integrate the equilibrium equations  (\ref{eqs1x}) for the
values (\ref{MWvals}), with $\A$ given by (\ref{Acase2}),
displaying the results in Fig. 2. Figures 2(a)--2(c) show how the
normalized potential, $\Psi$,\, the velocity squared and density
profiles, $\Vrot^2/\sigma_c^2$ and $Y$, are practically
indistinguishable from the curves of these same variables in the
case of an isothermal sphere coupled to a cosmological constant
(see \cite{SussHdez} for a comparison and a detailed discussion of
this system). As figure 2(a) shows, the function $\Psi$
oscillates, indicating ``turning points'' where $\Phi'=0$ that
separate regions with different signs of  $\Phi'$. The density
profile in figure 2(c) does not decay asymptotically as $1/r^2$,
as the isothermal profile, but oscillates around the density value
of the ``background'' $\Lambda$--field. At the first turning point
of $\Phi'$ the squared velocity $\Vrot^2$ (figure 2(b)) vanishes
and becomes negative (hence circular geodesic orbits cannot exist
for larger $r$). Notice that this first turning point occurs at
$r\approx 3.6$ \,Mpc, about an order of magnitude larger than the
physical radius of an isothermal halo similar to the Milky Way. As
discussed in \cite{SussHdez}, this fact allows one to ignore the
effects of a cosmological constant in the equilibrium of galactic
DM halos. A phenomenological analogue of a cosmological constant
can also be achieved by applying conditions (\ref{LambdaConds}) to
a force given by the coupling (\ref{Fab1}), though we have
preferred to illustrate this analogue with the case (\ref{Fab2})
because it offers a more natural association with a cosmic force
acting on very large length scales.
\\

\section{Discussion and conclusion}

We have derived the hydrostatic equilibrium conditions for a gas
with a specific class of self-interactions in a spherically
symmetric spacetime. This self-interaction is described by a
Lorentz--like force (\ref{force ansatz}) that is linear in the
particle momentum and characterized by a Maxwell--like
antisymmetric tensor $A^{ab}$. Within this framework, we have
considered two possible covariant forms for this force that are
adequate for the static and spherically symmetric spacetime
geometry: the one given in equation (\ref{Fab1}) where this tensor
is proportional to $\dot u^{[a}u^{b]}$, the other given in
equation (\ref{Fab2}), coupling it to the Riemann tensor. The
equilibrium equations in each case were presented in terms of
dimensionless variables,  which allowed us to obtain the Newtonian
type limit in a natural way by expanding with respect to the
dimensionless parameters $\xi$ and $\nu$, which characterize the
dispersion velocities associated with the hydrostatic ideal gas
pressure, $P=nk_BT$, and the pressure $\Pi$, respectively, the
latter being the result of the self--interaction on the
hydrodynamic level. We have provided two numerical examples of
Newtonian type weak field configurations, a modified isothermal
sphere and an example of how the curvature coupled force can act
phenomenologically as an effective cosmological constant.

While we do not claim the examples introduced in section IX and
depicted by figures 1 and 2 to describe ``realistic''
configurations, we believe that these simplest possible cases
nevertheless do illustrate how configurations of astrophysical
interest might arise from the presented formalism. In fact, these
models, especially that of the modified isothermal sphere shown in
figure 1, do convey interesting information worth discussing. For
example, numerical simulations~\cite{nbody_1, nbody_2, nbody_3,
urc, DMprofs1,DMprofs2,DMprofs3} yield halo configurations
characterized by empiric density and velocity profiles that differ
from the isothermal profiles, at least in the inner halo region
containing a disk of baryonic matter (which can be considered as
``tracers'' of the halo gravitational field). The non--isothermal
velocity profile associated with the Navarro--Frenk--White (NFW)
and analogous simulations~\cite{nbody_1,nbody_2,nbody_3}, as well
as other velocity profiles~\cite{urc,DMprofs1,DMprofs2,DMprofs3},
are qualitatively similar to the velocity profile shown by the
solid curve in Fig. 1(a), corresponding to an MB gas with a
generalized force of the type (\ref{Fab1}). Surprisingly, the
empiric density profiles of these simulations show a ``density
peak'' towards the halo center which does not match the density
profile of Fig. 1(c) that shows a ``flat density core''. Numerical
simulations also yield an asymptotic decay $\rho\propto1/r^3$,
that is different from the isothermal decay $\rho\propto1/r^2$ of
figure 1(c) (though it is difficult to verify the asymptotic DM
density decay by means of observations). However, observations in
low surface brightness (LSB) galaxies~\cite{vera,LSB} (supposedly
dominated by DM) do not reveal the ``density peak'' predicted by
numerical simulations, but an isothermal ``flat core'' (this is
still a hotly controversial issue). Hence, the simple toy model we
have proposed in section IX shows a rotation velocity profile
compatible with numerical simulations, but without the
controversial ``density peak''. It is then quite possible that
more elaborated models of a MB gas with self--interaction could
provide a reasonable phenomenological fit to observations,
especially if we consider the case with anisotropic stresses since
more realistic galactic halos are very unlikely to correspond to
isotropic configurations. Further research in this direction is
currently being undertaken.

It is expedient to point out that the interactions studied in this
paper seem to share features with MOND \cite{MOND}. This could
provide the theoretical basis for an approach in which the DM is
replaced by baryonic matter with a modified Newtonian limit of a
general relativistic description. In such a setting modifications
of Newton's law would be the result of a non-geodesic motion of
the (baryonic) particles of the medium.

 Finally, regarding the
phenomenological modelling of a cosmological constant by means of
the curvature coupled force (\ref{Fab2}), we have shown that this
self--interaction can reproduce similar effects as those of
previous studies of a cosmological constant (a cosmic $\Lambda$
field) in hydrostatic equilibrium with a MB gas (see
\cite{SussHdez}).  On the other hand, the gas dynamics with
self--interactions of the type that we have presented in this
paper has been successfully applied to dark energy models at the
cosmological scale~\cite{gasQ1,gasQ2,gasQ3,gasQ4}. Thus, the study
of the interplay and correspondence between galactic scale models,
as those derived here, and those describing the cosmic dynamics
dominated by dark energy provides an interesting and relevant
avenue for further research.
\\

\acknowledgements This work was supported by the Deutsche
Forschungsgemeinschaft, CONACYT (number E130.792), and by Russian
funds RFBR (grant N 04-05-64895) and HW (grant N 1789.2003.02).

\end{document}